\documentclass{article}
\usepackage{spconf,amsmath,graphicx}

\usepackage{amsfonts}
\usepackage{booktabs} 
\usepackage{tabularx}
\usepackage{tablefootnote}
\usepackage[symbol]{footmisc}

\usepackage{multirow} 
\usepackage[T1]{fontenc}
\usepackage{setspace}
\usepackage{threeparttable}
\usepackage[labelfont=bf]{caption} 
\usepackage{subcaption}
\usepackage[bookmarks=true,hypertexnames=true,pagebackref=true]{hyperref}
\hypersetup{
  colorlinks=true, 
  urlcolor=black,   
  linkcolor=blue,  
  citecolor=red    
}
\usepackage{color}
\usepackage[table]{xcolor}
\usepackage{cite}
\usepackage{circledsteps}
\usepackage{enumitem}

\makeatletter
  \newcommand{\linkdest}[1]{\Hy@raisedlink{\hypertarget{#1}{}}}
\makeatother
\newcounter{mathsymbolcount}
\newcommand{\mathsymbol}[1]{%
  \ifcsdef{symbol:#1}{%
    {\hypersetup{hidelinks}\hyperlink{symbol:#1}{#1}}%
  }{%
    \stepcounter{mathsymbolcount}%
    \linkdest{symbol:#1}{}#1%
    \global\expandafter\def\csname symbol:#1\endcsname{}
  }%
}

\newcolumntype{Y}{>{\centering\arraybackslash}X}

\renewenvironment{thebibliography}[1]{
  \begin{oldthebibliography}{#1}
  \setlength{\itemsep}{-0.1em}
  \setlength{\parskip}{0.em}
}
{
  \end{oldthebibliography}
}

\makeatletter
\def\bstctlcite{\@ifnextchar[{\@bstctlcite}{\@bstctlcite[@auxout]}}
\def\@bstctlcite[#1]#2{\@bsphack
  \@for\@citeb:=#2\do{%
    \edef\@citeb{\expandafter\@firstofone\@citeb}%
    \if@filesw\immediate\write\csname #1\endcsname{\string\citation{\@citeb}}\fi}%
  \@esphack}
\makeatother

\newcommand{\newpara}[1]{\vspace{2pt}\noindent\textbf{#1}}

\title{VoxtLM: Unified decoder-only models for consolidating speech recognition, synthesis and speech, text continuation tasks}
\name{Soumi Maiti$^1$, Yifan Peng$^1$, Shukjae Choi$^2$, Jee-weon Jung$^1$, Xuankai Chang$^1$, Shinji Watanabe$^1$}
\address{$^1$Carnegie Mellon University, USA \quad $^2$42dot Inc., Republic of Korea}

\begin{document}
\ninept
\bstctlcite{IEEEexample:BSTcontrol} 
\maketitle
\begin{abstract}
We propose a decoder-only language model, \textit{VoxtLM}, that can perform four tasks: speech recognition, speech synthesis, text generation, and speech continuation. 
VoxtLM integrates text vocabulary with discrete speech tokens from self-supervised speech features and uses special tokens to enable multitask learning. Compared to a single-task model, VoxtLM exhibits a significant improvement in speech synthesis, with improvements in both speech intelligibility from 28.9 to 5.6 and objective quality from 2.68 to 3.90.
VoxtLM also improves speech generation and speech recognition performance over the single-task counterpart. Further, VoxtLM is trained with publicly available data and training recipes and model checkpoints are open-sourced to make fully reproducible work. 
\end{abstract}

%

\section{Introduction}
\label{sec:intro}
\vspace{-3pt}
In recent years text language models (textLMs) have emerged as a powerful generative model in natural language processing (NLP)~\cite{radford2018gpt, radford2019gpt2, brown2020gpt3}. These textLMs can accommodate multiple tasks within a single model, leading to improvement in performance across a variety of tasks. On the other hand, with advances in discrete speech representations speech language models (speechLMs)~\cite{lakhotia2021generative, borsos2022audiolm} have also been proposed. However, prior speechLMs focus on individual tasks, such as speech continuation or text-to-speech (TTS)~\cite{hayashi2020discretalk, wang2023valle}. Our hypothesis is that by unifying diverse speech tasks into a generative language model (LM), we can potentially address multiple speech tasks using a single model with improved generalization thanks to multitask learning.

\begin{figure}[!t]
\begin{minipage}[b]{0.9\linewidth}
  \centering
  \centerline{\includegraphics[width=0.8\linewidth, trim={0.38cm 0.65cm 0.65cm 0.3cm},clip]{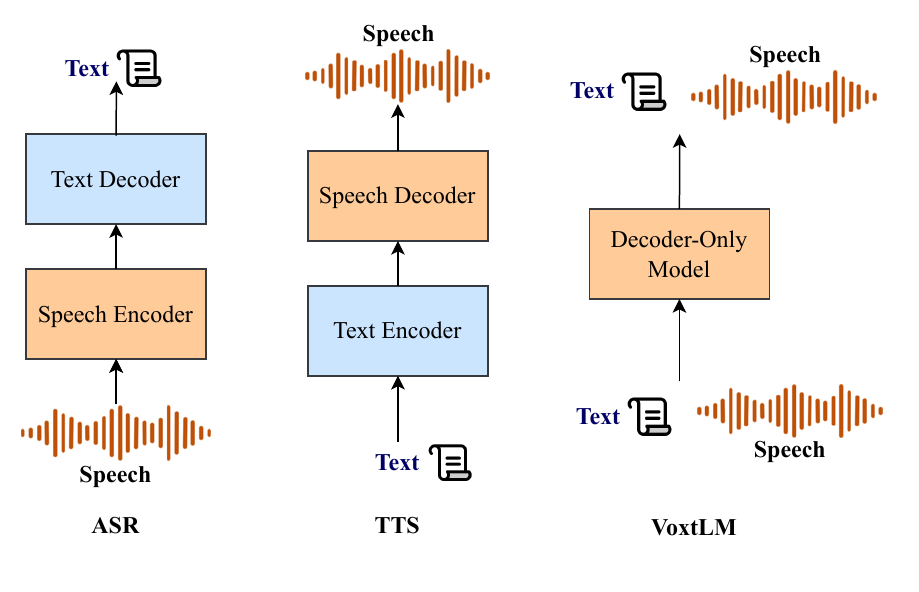}}
\end{minipage}
\caption{ASR and TTS use encoder-decoder architecture while VoxtLM is decoder-only. In VoxtLM, all parameters are shared between speech and text modalities, compared to separate encoder/ decoder for speech and text.}
\label{fig:arch}
\end{figure}

Traditionally, speech applications such as automatic speech recognition (ASR) and text-to-speech (TTS) use encoder-decoder architectures~\cite{prabhavalkar2023end, shen2018tacotron2, ren2020fastspeech2}. These architectures consist of an encoder, for input processing and a decoder, for generating the output. For example, speech-to-text involves a speech encoder and a text decoder, whereas text-to-speech employs a text encoder and a speech decoder. Integrating task-specific and modality-specific encoder-decoder components complicates the incorporation of multiple tasks~\cite{ao-etal-2022-speecht5, chen2022maestro}. In contrast, we can simplify multitask integration with a joint speech-text decoder-only model (depicted in Fig. ~\ref{fig:arch}).

In this work, we investigate two main questions. Firstly, can we cast diverse speech tasks as language modeling? ASR and TTS are used as example speech tasks. Secondly, can we combine speech tasks in a joint speech-text language modeling framework?
To this purpose, we introduce a novel LM framework \textit{VoxtLM} (\textbf{Vo}ice-te\textbf{xt} \textbf{L}anguage \textbf{M}odel). VoxtLM combines multiple speech tasks within a single autoregressive decoder model. Specifically, we combine four tasks: speech recognition (speech-to-text), speech synthesis (text-to-speech), text generation (text-to-text), and speech generation (speech-to-speech). We create a \textit{Voxt} (voice + text) \textit{vocabulary} by merging self-supervised discrete speech tokens with the text vocabulary and incorporate sub-word modeling to efficiently process long sequences of speech.
We show that VoxtLM can model both ASR and TTS as conditioned language model. In addition, combining four tasks leads to improvement in speech generation, ASR, and TTS. Significant improvement is observed in the TTS task with improvement in both intelligibility (28.9 to 5.6) and neural-predicted quality (2.68 to 3.90). Additionally, we demonstrate that improved initialization with pretrained textLM and scaling model parameters help in ASR. To ensure reproducibility, we use publicly available datasets, open-source our training and inference in the form of open-source toolkit ESPnet recipe\footnote{\scriptsize{\url{https://github.com/ESPnet/ESPnet}}} and make model checkpoints available. TTS samples are also available.\footnote{\scriptsize\url{https://soumimaiti.github.io/icassp24_voxtlm/}}


\begin{table*}[!tb]
\caption{\textit{Voxt} data format for different tasks: training and inference. Inference: provided conditions for generating prediction.}
\label{tab:format}
\centering
\resizebox{0.8\linewidth}{!}{
\begin{tabular}{ l | l | l | c }
\toprule
Task &  \multicolumn{1}{c|}{Training} & \multicolumn{2}{c}{Inference}  \\
 &  & \multicolumn{1}{c}{Condition} & Prediction \\
\midrule 
TextLM & $\langle$\texttt{generate-text}$\rangle, Y$  &
$\langle\texttt{generate-text}\rangle , Y^{\text{test}}$ & $\hat{Y}$ \\
SpeechLM & $\langle$\texttt{generate-speech}$\rangle, D $ & $\langle\texttt{generate-speech}\rangle ,D^{\text{test}}$ & $\hat{D}$ \\
ASR & $\langle$\texttt{start-speech}$\rangle$, $D$, $\langle$\texttt{generate-text}$\rangle$, $Y$ & $\langle\texttt{start-speech}\rangle ,D^{\text{test}},\langle\texttt{generate-text}\rangle$ & $\hat{Y}$ \\
TTS & $\langle$\texttt{start-text}$\rangle, Y, \langle$\texttt{generate-speech}$\rangle, D$ & $\langle\texttt{start-text}\rangle,Y^{\text{test}},\langle\texttt{generate-speech}\rangle$ & $\hat{D}$ \\
\bottomrule
\end{tabular}
}
\end{table*}

\vspace{-3pt}
\section{Related Work}
\label{sec:rel_work}
\newpara{Discrete speech representations.} Speech signals can be represented as two types of discrete tokens: semantic tokens and acoustic tokens. Semantic tokens are quantized from self-supervised learning features (e.g., HuBERT~\cite{hsu2021hubert}, w2v-BERT~\cite{Chung2021w2vbert}) through clustering, which mostly captures the linguistic content. Acoustic tokens are generated by audio codec models~\cite{Zeghidour2022soundstream, defossez2022highfi}. They capture rich acoustic information which is suitable for high-quality speech synthesis, but they consist of multiple code streams and are thus difficult to model. In this work, we follow GSLM~\cite{lakhotia2021generative} to use semantic tokens derived from HuBERT.

\newpara{Joint modeling of speech and text.} Several studies~\cite{ao-etal-2022-speecht5, bapna2021slam, chen2022maestro} propose to learn shared speech-text representations in a self-supervised manner. However, they employ separate encoders and decoders for different modalities. They also require additional losses like an alignment loss to encourage cross-modal transfer between speech and text. Recent concurrent studies employ a single model for multiple speech and text conversion tasks~\cite{dong2023polyvoice, wang2023viola, zhang2023speechgpt}, which are similar to our approach.
SpeechGPT~\cite{zhang2023speechgpt} uses a three-stage adaptation to combine audio generation with textLMs. PolyVoice~\cite{dong2023polyvoice} applies speechLM to speech-to-speech translation (S2ST) with three decoder-only LMs. VioLA~\cite{wang2023viola} extends VALL-E~\cite{wang2023valle} for ASR and S2ST. Among them, VioLA is the most related method to this work. However, VioLA does not incorporate speech or text continuation tasks and requires additional sequence modeling for speech representations, which makes it more complicated than our approach. Moreover, we utilize textually pre-trained OPT~\cite{zhang2022opt} for better initialization inspired by \cite{hassid2023twist} and leverage different speech tokens. Also in comparison to other works, our work is fully reproducible. 

\begin{figure}[!t]
\begin{minipage}[b]{1.0\linewidth}
  \centering
  \centerline{\includegraphics[width=1.0\linewidth, trim={0.5cm 0cm 0cm 0cm},clip]{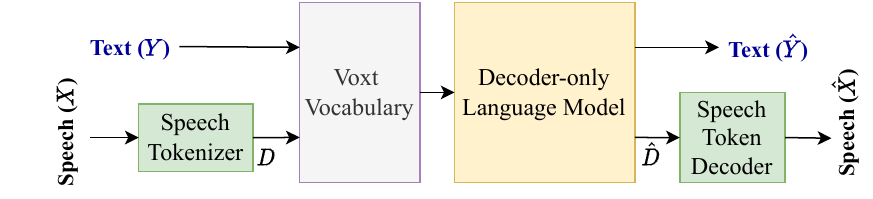}}
\end{minipage}\vspace{3pt}
\caption{Overview of VoxtLM, our proposed autoregressive decoder-only LM incorporating speech and text within an integrated vocabulary $\mathcal{V}_{\text{voxt}}$. The model uses two additional modules, the speech tokenizer and the speech token decoder to facilitate the conversion between continuous speech signal and discrete speech tokens.}
\label{fig:model}
\end{figure}


\section{Method}
\label{sec:method}
\vspace{-3pt}
Consider $Y=(y_i \in \mathcal{V}_{\text{txt}}| i= 1, \cdots, {t_{\text{txt}}})$ is a text utterance from a vocabulary $\mathcal{V}_{\text{txt}}$ with length $t_{\text{txt}}$. The probability of $Y$ can be expressed as $ p(Y)= \Pi_{i=1}^{t_{\text{txt}}} p(y_i| y_1, \cdots, y_{i-1}).$
Now, when dealing with a continuous speech signal, we can convert it into discrete speech tokens (dst), represented as $D = (d_i \in \mathcal{V}_{\text{dst}}| i=1, \cdots, t_{\text{dst}} )$ using a tokenizer. In this context $\mathcal{V}_{\text{dst}}$ is the vocabulary of discrete speech tokens. These discrete speech tokens can be treated as spoken language within $\mathcal{V}_{\text{dst}}$ and modeled in a manner similar to text.
We combine text and speech in a new vocabulary \textit{Voxt vocabulary} by  $\mathcal{V}_{\text{voxt}}=\mathcal{V}_{\text{txt}} \cup \mathcal{V}_{\text{dst}}$. Therefore, we can model the probability of both speech and text tokens as $Z$, where $Z=( z_i \in \mathcal{V}| i=1, \cdots, t)$. This probability is expressed as:
\begin{equation}\label{eq:auto_z}
    p(Z)= \Pi_{i=1}^{t} p(z_i| z_1, \cdots, z_{i-1}).
\end{equation}
Here, $Z$ can represent discrete speech tokens $D(\mathcal{V}=\mathcal{V}_{\text{dst}}) $ or text tokens $Y (\mathcal{V}=\mathcal{V}_{\text{txt}})$ or various combinations of $Y$ and $D$.  

\vspace{-5pt}
\subsection{VoxtLM}\label{sec:model}
Fig.~\ref{fig:model} illustrates the model's overall architecture. Input of VoxtLM can be both speech and text within the $\mathcal{V}_{\text{voxt}}$ vocabulary. To process speech, we use two additional modules to convert between continuous and discrete domains in speech. The speech tokenizer maps $X$ to $D$, while the speech token decoder maps generated $\hat{D}$ back to $\hat{X}$. Similar to~\cite{lakhotia2021generative}, our speech tokenizer uses $k$-means clustering to derive discrete features from the pretrained HuBERT~\cite{hsu2021hubert}. It is worth noting that selecting a small $k$ value may capture linguistic information effectively, but might fall short in representing other acoustic aspects particularly crucial for speech synthesis. We experiment with different $k$ to assess the impact.
Furthermore, within $\mathcal{V}_{\text{voxt}}$ vocabulary, we apply subword modeling~\cite{kudo2018sentencepiece, sennrich2016neural, kudo2018subword} to replace frequent patterns with metatokens. Such subword modeling technique is used to include more contextual information in text~\cite{radford2018gpt} or to reduce the long sequence length of speech~\cite{chang2023discreteasr}.

\vspace{-8pt}
\subsubsection{Data format}\label{ssec:data_format}
\vspace{-3pt}
We use special tokens to guide the model in performing various tasks.
Four such tokens are used: $\langle$\texttt{start-text}$\rangle$ and $\langle$\texttt{start-speech}$\rangle$ indicate the beginning of text or speech conditioning in the language model. $\langle$\texttt{generate-speech}$\rangle$ and $\langle$\texttt{generate-text}$\rangle$ instruct the model whether to generate speech or text.
Table~\ref{tab:format} shows examples of the \textit{Voxt data format} for various tasks during training. Ideally, we can extend to more tasks with additional task-specific tokens.

\vspace{-5pt}
\subsubsection{Training}\label{ssec:training}
\vspace{-3pt}
VoxtLM consists of an embedding layer and a series of transformer~\cite{vaswani2017transformer} decoder layers. The embedding layer maps input $Z$ (in Eq. \ref{eq:auto_z}) into $F$-dimensional feature space, $E=(e_i \in \mathbb{R}^F|i=1, \cdots, t)$ using an embedding table of size $|\mathcal{V}_{\text{voxt}}|\times F$. We use $L$ transformer decoder layers with $H$ attention heads. The model's output includes a linear layer followed by softmax, generating a probability distribution over the tokens in $\mathcal{V}_{\text{voxt}}$. VoxtLM is trained as an autoregressive language model. 
In training, teacher forcing is used for the preceding tokens. Given $Z$, at each timestep $i$, predicted distribution is $\hat{p}_i=\text{VoxtLM}(z_1,\cdots, z_{i-1})$. Given true probability distribution $p_i$, the loss is calculated using cross-entropy as
$L_{\text{CE}}(p_i, \hat{p}_i) = - \sum_{c=1}^{\mathcal{V}_{\text{voxt}}} p_i(c) \log \hat{p}_i(c)$.


\begin{table}[tb]
\caption{Number of utterances used in training of different VoxtLM setups. Bal: balanced data for four tasks; and 3M: uses the same number (3M) of text-only and speech-only utterances, a balanced setup for total text and total speech data.}
\label{tab:utt}
\centering
\scalebox{0.8}{
\footnotesize
\begin{tabular}{ l | c c| c c| c }
\toprule
\multirow{2}{*}{Source} & \multicolumn{2}{c|}{\textbf{Unpaired}} & \multicolumn{2}{c|}{\textbf{Paired}} & \multirow{2}{*}{Configuration} \\
 & Speech & Text & ASR & TTS \\
\midrule
$\mathcal{D}_{\text{Bal}}$ & 300K & 300K & 281K & 404K & LL+LS+LT+VC \\
$\mathcal{D}_{\text{3M}}$ & ~~~~3M & ~~~~3M  & 281K & 404K & LL+LS+LT+VC \\
$\mathcal{D}_{\text{Set}}$ & ~~12M &  ~~40M & 281K & 404K & LL+LS+LT+VC\\
$\mathcal{D}_{\text{Set+}}$ & ~~12M & ~~40M & ~~11M & 404K & LL+LS+LT+VC+MLS\\
\bottomrule
\end{tabular}
}
\end{table}

\begin{table*}[!tb]
\caption{Experimental results comparing multitasking VoxtLM against \textit{four} single-task VoxtLM for textLM, speechLM, ASR and TTS. We use token size ($k$) 50 for all models. Single-task models are trained with all available data, for VoxtLM we report different training data (Table~\ref{tab:utt}) cases . For ASR we report test-clean/test-other results. $\mathcal{D}_{\text{Set}}^{*}$: four single-task models whereas other rows depict multitask model.}

\label{tab:multitask}
\centering
\resizebox{0.8\linewidth}{!}{
\footnotesize
\begin{tabular}{ l | c | c c c | c c c | c | c c }
\toprule 
\multirow{2}{*}{Source} & \multirow{2}{*}{\# params} &  \multicolumn{3}{c|}{\textbf{TextLM}} & \multicolumn{3}{c|}{\textbf{SpeechLM}} & {\textbf{ASR}} & \multicolumn{2}{c}{\textbf{TTS}} \\
 &  & PPL($\downarrow$) & sWUGGY($\uparrow$) & sBLIMP($\uparrow$) & PPL($\downarrow$) & sWUGGY($\uparrow$) & sBLIMP($\uparrow$) & WER($\downarrow$) & CER($\downarrow$) & MOSNet($\uparrow$)\\
\midrule
$\mathcal{D}_{\text{Set}}^{*}$ &125M & 18.3 & 77.1 & \textbf{80.3} & 73.8 & 62.9 & 53.9 & ~~8.8 / 21.4 & 28.9 & 2.68\\
\midrule
$\mathcal{D}_{\text{Bal}}$ & 125M & 15.4 & 77.7 & 66.7 & 68.5 & 60.7 & 52.7 & ~~\textbf{8.6} / \textbf{20.9} &  ~~\textbf{5.6} & 3.76\\
$\mathcal{D}_{\text{3M}}$ & 125M & 13.5 & 77.9 & 68.0 & 58.1 & \textbf{63.6} & \textbf{55.2} & 11.0 / 24.4 & ~~7.0 & \textbf{3.90} \\
$\mathcal{D}_{\text{Set}}$ & 125M & 11.1 & \textbf{80.3} & 74.2 & 62.1 & 62.8 & 54.1 & 21.0 / 37.4  & ~~8.8 & 3.86\\
\bottomrule
\end{tabular}
}
\vspace{-.4cm}
\end{table*}

\vspace{0.2cm}
\newpara{Initialization with pretrained textLM}.
Previous work ~\cite{hassid2023twist} shows that in speechLM initializing with a pre-trained textLM achieves better performance and faster convergence. Motivated by this, we use the pretrained textLM OPT~\cite{zhang2022opt} to initialize VoxtLM weights and learn the embedding table from scratch. The same model configuration is used as the pretrained model except for $|\mathcal{V}_{\rm voxt}|$. OPT is used due to training on publicly available data and the availability of smaller pretrained models.

\vspace{-5pt}
\subsubsection{Inference}\label{ssec:inference}
\vspace{-3pt}
The prediction from VoxtLM is expressed as:
\begin{equation}
    \text{prediction} \leftarrow p(\cdot | \text{condition}).
\end{equation}
For TTS, condition is the test text utterance $Y^{\text{test}}$ and prediction is speech tokens $\hat{D}$. In ASR, condition is test speech tokens $D^{\text{test}}$ and prediction is the recognized text $\hat{Y}$. For speech continuation, condition involves prefix speech tokens $D^{\text{test}}$ and prediction is continued speech tokens $\hat{D}$. For text continuation, the condition is text $Y^{\text{test}}$ and prediction is continued text $\hat{Y}$ (summarized in Table~\ref{tab:format}.
We use beam search in the inference phase.

\newpara{Speech token decoder.}
The speech token decoder takes both $\hat{D}$ and a speaker embedding $s_{\text{spk}} \in \mathbb{R}^N$ of dimensionality $N$ as inputs and produces $\hat{X}$.
We use the HiFiGAN~\cite{Kong2020hifigan} as the architecture and x-vector~\cite{xvector8461375} as speaker embedding vector.

\subsection{Evaluation Metrics}
\vspace{-3pt}

\begin{itemize}[leftmargin=8pt]\setlength\itemsep{0em}
\item For speech and text generation, we use perplexity (PPL) for evaluating models with same vocabulary size. For different vocabulary size models, we use spot-the-word error using sWUGGY and syntactic score using sBLIMP dev set~\cite{nguyen2020zero}. sWUGGY and sBLIMP are chosen as other speech LM works also report them.  
\item For ASR, we use the word error rate (WER).
\item For TTS, we measure intelligibility with character error rate (CER) and quality using the neural-predicted mean opinion score (MOS) with MOSNet~\cite{lo2019mosnet, cooper2022generalization}. We choose neural MOS prediction model because it scales to large number of evaluations and shows high-correlation with TTS evaluations in English.  
\end{itemize}


\section{Experiments}
\newpara{Dataset.}
We use a combination of speech-only, text-only, and paired speech-text datasets from public corpora. 
\begin{itemize}[leftmargin=5pt]\setlength\itemsep{0em}
\item \textit{Speech-only data}: we use LibriLight (LL)~\cite{Kahn2020librilight} with 60K hours of audiobook speech from 7K speakers (12M utterances). 
\item \textit{Text-only data}: we use the Librispeech (LS)~\cite{panayotov2015librispeech} external textLM dataset (40M text utterances). \item \textit{Speech-text paired data}: 
\begin{itemize}[leftmargin=3pt]\setlength\itemsep{0em}\vspace{-3pt}
\item For ASR, we use Librispeech~\cite{panayotov2015librispeech} with 960 hours of data ( 281K utterances). 
For an additional supervised data experiment, we use English Multilingual Librispeech (MLS)~\cite{pratap2020mls} with 44K hours of data from 5490 speakers (11M utterances). 
\item For TTS, we use LibriTTS (LT)~\cite{zen2019libritts} with 580 hours of audiobook data from 2456 speakers and VCTK (VC)~\cite{veaux2017vctk} with 44 hours of studio recorded data from 109 speakers (404K utterances). 
\end{itemize}\vspace{-3pt}
\end{itemize}
We standardized the data by downsampling speech to a 16kHz rate, converting text to lowercase, and removing punctuation.
We use separate test/dev sets for each task. For textLM and speechLM, we use the test set from LS and dev sets from sWUGGY and sBLIMP, \textit{text} for textLM, and \textit{speech} counterpart for speechLM. For ASR we use \textit{speech-text} test set from LS test-clean and test-other and report both test-clean/test-other separately.
In TTS for computational efficiency, we create a test set of 100 utterances from two speakers from the LT test-clean. The test speakers are chosen via random sampling (specifically, speaker ids 1089 and 1284).

\newpara{Experimental setup.} To train the sub-word model, we use paired text-speech from ASR and TTS datasets. We experiment with three $k$ values (introduced in Sec.~\ref{sec:model}), 50, 200, and 1000, denoted as \textit{VoxtLM}-$k$. We also vary BPE sizes, setting them at $2\text{K}$, $5\text{K}$, and $10\text{K}$ for $k$ values 50, 100 and 200, respectively. 
We use three configurations, small ($L$=12, $F$=768, $H$=12), medium ($L$=24, $F$=1024, $H$=16), and large ($L$=24, $F$=2048, $H$=32), with $L$, $H$ and $F$ detailed in Sec.~\ref{ssec:training}. We use 4 A100 GPUs for training small/medium and 8 A100 GPUs for large with Adam optimizer~\cite{kingma2014adam} and warmup learning rate schedule.
Training data size varies considerably between different tasks. For example, the paired data for ASR and TTS are $100\times$ smaller than text-only data and $40\times$ smaller than speech-only data. We can assume that achieving optimal performance across all tasks requires balanced data for each of them. It is also worth noting that text-only data is more readily available compared to speech-only and paired data. Nonetheless, to assess the effect of different dataset sizes for tasks, we consider balanced and unbalanced data sets for training, as summarized in Table~\ref{tab:utt}.


\begin{table}[tb]
\caption{ Experimental results comparing with and without initialization with pretrained (PT) textLM for VoxtLM-$k$50 with $\mathcal{D}_\text{Set}$.}
\label{tab:textlm_init}
\centering
\resizebox{1\linewidth}{!}{
\begin{tabular}{ l | c | c | c | c | c c }
\toprule 
\multirow{2}{*}{Name} & \multirow{2}{*}{\begin{tabular}[c]{@{}c@{}}\# params\end{tabular}} & \textbf{TextLM} & \textbf{SpeechLM} & \textbf{ASR} & \multicolumn{2}{c}{\textbf{TTS}} \\
 &  & PPL($\downarrow$) & PPL($\downarrow$) & WER($\downarrow$) & CER($\downarrow$) & MOSNet($\uparrow$) \\
\midrule
w/o PT & 125M & 11.1 & 62.1 & 21.0 / 37.4 & ~~\textbf{8.8} & 3.86 \\
w/ ~~PT & 125M  & \textbf{10.4} & \textbf{58.8} & \textbf{13.1} / \textbf{28.8} & ~~9.4 & \textbf{3.92} \\
\bottomrule
\end{tabular}
}
\end{table}

\begin{table*}[tb]
\caption{Experimental results comparing speech token size $k$ for VoxtLM. We compare the two conditions: $\mathcal{D}_\text{Bal}$ and $\mathcal{D}_\text{Set}$ (Table~\ref{tab:utt}). $^\dagger$ denotes initialization with OPT.}
\label{tab:discrete}
\centering
\resizebox{0.85\linewidth}{!}{
\footnotesize
\begin{tabular}{ l | c | c | c c c | c c c | c | c c }
\toprule 
\multirow{2}{*}{Name} & \multirow{2}{*}{Source} &  \multirow{2}{*}{\begin{tabular}[c]{@{}c@{}}\# params\end{tabular}} & \multicolumn{3}{c|}{\textbf{TextLM}} & \multicolumn{3}{c|}{\textbf{SpeechLM}} & {\textbf{ASR}} & \multicolumn{2}{c}{\textbf{TTS}} \\
 &  &  &  PPL($\downarrow$) & sWUGGY($\uparrow$) & sBLIMP($\uparrow$) & PPL($\downarrow$) & sWUGGY($\uparrow$) & sBLIMP($\uparrow$) & WER($\downarrow$) & CER($\downarrow$) & MOSNet($\uparrow$)\\
\midrule
VoxtLM-$k$50 & $\mathcal{D}_\text{Bal}$ &125M & 15.4 & \textbf{77.7} & 66.7 & 68.5 & 60.7 & \textbf{52.7} & ~8.6 / 20.9 & 5.6 & 3.76\\
VoxtLM-$k$200  & $\mathcal{D}_\text{Bal}$ & 125M & 21.6 & 77.3 & \textbf{67.9} & 58.6 & \textbf{61.6} & 52.1 & ~6.1 / 15.4 & 3.2 & \textbf{4.36}\\
VoxtLM-$k$1000 & $\mathcal{D}_\text{Bal}$& 125M & 26.3 & 76.4 & 67.6 & 38.7 & 60.7 & 52.5 & ~\textbf{5.4} / \textbf{14.5} & \textbf{2.6} & 4.30  \\
\midrule
VoxtLM-$k$50$^{\dagger}$ & $\mathcal{D}_\text{Set}$& 350M & 10.3 & \textbf{81.0} & 75.1 & 68.2 & 62.7 & 53.8 & 13.5 / 27.2 & 6.6 & 3.91 \\
VoxtLM-$k$200$^{\dagger}$ &$\mathcal{D}_\text{Set}$&  350M & 12.7 & 80.2 & \textbf{78.8} & 45.7 & \textbf{65.5} & \textbf{55.3} & ~~\textbf{6.5} / \textbf{17.6} & \textbf{3.5} & \textbf{4.36} \\ 
\bottomrule
\end{tabular}
}
\end{table*}
\begin{table*}[tb]
\caption{Experimental results comparing larger model size and more supervised data for VoxtLM. $^\dagger$ denotes initialization with OPT.}
\label{tab:scalability}
\centering
\resizebox{0.85\linewidth}{!}{
\footnotesize
\begin{tabular}{ l | l | c | c c c | c c c | c | c c }
\toprule 
\multirow{2}{*}{Name} & \multirow{2}{*}{Source} & \multirow{2}{*}{\begin{tabular}[c]{@{}c@{}}\# params\end{tabular}} & \multicolumn{3}{c|}{\textbf{TextLM}} & \multicolumn{3}{c|}{\textbf{SpeechLM}} & {\textbf{ASR}} & \multicolumn{2}{c}{\textbf{TTS}} \\
 &  &  &  PPL($\downarrow$) & sWUGGY($\uparrow$) & sBLIMP($\uparrow$) & PPL($\downarrow$) & sWUGGY($\uparrow$) & sBLIMP($\uparrow$) & WER($\downarrow$) & CER($\downarrow$) & MOSNet($\uparrow$)\\
\midrule
VoxtLM-$k$200$^{\dagger}$ & $\mathcal{D}_\text{Set}$\:  &350M & 12.7 & 80.2 & 78.8 & 45.7 & 65.5 & 55.3 & ~6.5 / 17.6 & \textbf{3.5} & \textbf{4.36} \\ 
VoxtLM-$k$200$^{\dagger}$ & $\mathcal{D}_\text{Set}$\: & \;1.3B  & \textbf{11.4} & 80.9 & 80.2 & 42.1 & \textbf{66.1} & 56.7 & ~4.6 / 12.1 & 3.9 & 4.33 
\\
VoxtLM-$k$200$^{\dagger}$ & $\mathcal{D}_\text{Set+}$ & 350M & 13.6 & 80.1 & 77.0 & 45.5 & 64.1 & 55.0 & ~3.5 / ~~8.7 & 3.5 & 4.33 \\
VoxtLM-$k$200$^{\dagger}$ & $\mathcal{D}_\text{Set+}$ & \;1.3B & 12.1 & \textbf{81.2} & 80.0 & \textbf{40.9} & 65.6 & \textbf{57.1} & ~\textbf{2.7} / ~~\textbf{6.5} & 3.6 & 4.33 \\
\midrule
VoxtLM-$k$1000$^{\dagger}$ & $\mathcal{D}_\text{Set+}$ & 125M & 18.3 & 77.3 & 75.2 & 36.7 & 60.7 & 51.3 & ~4.7 / 11.9 & 3.9 & 4.28~ \\
VoxtLM-$k$1000$^{\dagger}$ & $\mathcal{D}_\text{Set+}$ &350M & 16.4 & 79.6 & 77.8 & 32.3 & 62.1 & 53.2 & ~3.5 / ~~8.6 & 6.1 & 4.27 \\
\bottomrule
\end{tabular}
}
\end{table*}

\begin{table}[tb]
\caption{SpeechLM and ASR results: Comparison with the state-of-the-art models with VoxtLM. $^\dagger$ denotes initialization with OPT.}
\label{tab:speechlm_asr}
\centering
\resizebox{1.0\linewidth}{!}{
\footnotesize
\begin{tabular}{ l | l | c | c c | c }
\toprule
\multirow{2}{*}{Name} & \multirow{2}{*}{Source} &  \multirow{2}{*}{\begin{tabular}[c]{@{}c@{}}\# params\end{tabular}}  & \multicolumn{2}{c|}{\textbf{SpeechLM}} & {\textbf{ASR}} \\
 &  & & sWUGGY($\uparrow$) & sBLIMP($\uparrow$) & WER($\downarrow$)\\
\midrule
GSLM-$k$50~\cite{lakhotia2021generative} &  ~~~~LL & 172M  & - & 55.9 &  -~~ /~~ - \\
GSLM-$k$200~\cite{lakhotia2021generative} & ~~~~LL & 172M & - & 53.0 & - ~~/~~ - \\
AudioLM~\cite{borsos2022audiolm} &  ~~~~LL &  900M &71.5 & 64.7 & - ~~/~~ -\\
ASR-Fbank & ~~~~LS & 149M & - & - & ~2.2 / ~~4.6\\
dst-ASR-Hubert &  ~~~~LS & ~~39M & - & - & ~4.2 / 10.8 \\
\midrule
VoxtLM-$k$200$^{\dagger}$  & ~$\mathcal{D}_\text{Set}~~~\:$ & \;1.3B & 66.1 & 56.7 & ~4.6 / 12.1 \\ 
VoxtLM-$k$200$^{\dagger}$ & ~$\mathcal{D}_\text{Set+}$ & \;1.3B & 65.6 & 57.1 & ~2.7 / ~~6.5 \\
\bottomrule
\end{tabular}
}
\end{table}


\begin{table}[tb]
\footnotesize
\caption{ Comparison with the state-of-the-art baseline for TTS with VoxtLM. $^\dagger$ denotes initialization with OPT. }
\label{tab:textlm_tts}
\centering
\resizebox{0.9\linewidth}{!}{
\footnotesize
\begin{tabular}{ l | c | c | c c }
\toprule
\multirow{2}{*}{Name} & \multirow{2}{*}{Source} & \multirow{2}{*}{\begin{tabular}[c]{@{}c@{}}\# params \end{tabular}} & \multicolumn{2}{c}{\textbf{TTS}} \\
 & & & CER($\downarrow$) & MOSNet($\uparrow$)\\
\midrule
VITS~\cite{vitskim2021conditional} & LT & ~~~97M & 7.7 & 4.20 \\
\midrule
VoxtLM-$k$200$^{\dagger}$ & $\mathcal{D}_\text{Set}$\,& ~350M &  3.5 & \textbf{4.36} \\
VoxtLM-$k$1000 & $\mathcal{D}_\text{Bal}$ &   ~125M & \textbf{2.6} & 4.30  \\
\bottomrule
\end{tabular}
}
\end{table}


\subsection{Results}
\vspace{-3pt}
\newpara{Single vs multitask.} We compare multitask and \textit{four} single-task models using VoxtLM-$k$50. Single-task LMs are trained separately for each task (ASR, TTS, speechLM, and textLM) and are reported in the first row of Table~\ref{tab:multitask}, with each column representing a separate single-task model. Compared to single-task, VoxtLM shows competitive results for all four tasks, although the best model differs. For textLM $\mathcal{D}_\text{Set}$ exhibits higher sWUGGY but lower sBLIMP score. In speechLM, $\mathcal{D}_\text{3M}$ has the best scores in both sWUGGY and sBLIMP, followed by $\mathcal{D}_\text{Set}$. In TTS, all multitask models show improvement compared to single task. ASR reports improvement in $\mathcal{D}_\text{Bal}$. We note that ASR is most affected in the unbalanced case: probably due to the lower ASR data ratio to textLM/speechLM ($100\times$/$40\times$ less). A smaller degradation in ASR is also observed in $\mathcal{D}_\text{3M}$ where ASR data ratio to textLM/speechLM is relatively better ($10\times$ less). 

\newpara{Initialization with pretrained textLM.}  
We compare with and without initialization with OPT for VoxtLM-$k$50 with $\mathcal{D}_\text{Set}$ and report in Table~\ref{tab:textlm_init}. 
Initialization improves the performance of three tasks: textLM, speechLM, and ASR. For TTS, a slight degradation in CER is observed, whereas objective quality improves. In particular, better initialization aids ASR performance in the unbalanced scenario (reducing test-clean WER from 21.0 to 13.1).

\newpara{Effect of token vocabulary size.} We compare $k$=50, 
 200 and 1000 as outlined in Table~\ref{tab:discrete}. Comparisons are made in $\mathcal{D}_\text{Bal}$ and $\mathcal{D}_\text{Set}$. For ASR and TTS, performance of $k$=50 is poor. For speechLM with $\mathcal{D}_\text{Set}$ best sores on sWUGGY and sBLIMP are observed with the $k$=200 model. TextLM, as expected, does not show a significant pattern with varying $k$. 

\newpara{Scalability.} Next, we explore whether model size can help with data balancing by comparing medium and large models with $k$=200, presented in Table~\ref{tab:scalability}. All metrics in TextLM, speechLM, and ASR show improvement with larger model. TTS shows a very small degradation in intelligibility ($0.4$) and quality ($0.03$). To mitigate the smaller ratio of paired data, we incorporate more supervised data for ASR in $\mathcal{D}_\text{Set+}$. We compare with $k$ = 200 and $k$ = 1000 and observe an improvement in the ASR task.

\newpara{Comparison with single-task state-of-the-arts.} 
Furthermore, we compare with state-of-the-art models in TTS, ASR, and speechLM. Note that these models aren't fully comparable due to differences in training data, strategies, and architecture. Following models are used: for speechLM, we use GSLM~\cite{lakhotia2021generative} and AudioLM~\cite{borsos2022audiolm}, for TTS, we use VITS~\cite{vitskim2021conditional} and for ASR we use E-Branchformer~\cite{kim2023branchformer}. For ASR, we compare two models: one using spectrogram as input (ASR-Fbank) and another using discrete speech tokens as input (dst-ASR-Hubert), trained following the procedure~\cite{chang2023discreteasr} and the same speech tokenizer as VoxtLM-$k$1000. We use a pretrained VITS model with LibriTTS. For speechLM (Table~\ref{tab:speechlm_asr}), GSLM-$k$200 which uses the same tokenizer and a similar one-stage model, sBLIMP score is lower compared to VoxtLM. However, in AudioLM which uses two token representations (acoustic and semantic) and a three-stage model, both sWUGGY and sBLIMP scores are higher, suggesting potential for further improvement with hierarchical tokens and multistage training. For ASR, compared to dst-ASR-Hubert, which used the same tokenizer as VoxtLM, we observe a lower WER. Compared to ASR-Fbank (no tokenizer), WER is higher, such a trend is also observed in other discrete ASR models~\cite{chang2023discreteasr}.
In TTS (Table~\ref{tab:textlm_tts}), compared to VITS, VoxtLM reports better intelligibility and quality. Although VoxtLM is trained with a larger data set compared to VITS, it is interesting to note that for traditional TTS diverse training data with more noise and more speakers degrade performance but here improvement is observed.

Finally, our experimental results show that both ASR and TTS can be modeled as language modeling tasks. Moreover, using special tokens we can combine ASR and TTS with joint speech-text language modeling framework. Although the four tasks are quite different, combining four tasks leads to improvement.


\section{Conclusion}
\vspace{-5pt}
The integration of speech and text tasks within a joint language modeling framework presents a promising avenue for speech processing. We present a special token-based approach to combine four speech and text tasks: speech recognition, speech synthesis, text and speech generation. Our results demonstrate that by integrating different speech tasks into one generative model, we can improve the performance of the tasks. In particular, TTS shows impressive performance compared to the state-of-the-art VITS. We will expand this work to include more speech tasks in the future.

\newpara{Acknowledgements}
\footnotesize
Experiments of this work used the Bridges2 system at PSC and Delta system at NCSA through allocations CIS210014 and IRI120008P from the Advanced Cyberinfrastructure Coordination Ecosystem: Services \& Support (ACCESS) program, supported by National Science Foundation grants \#2138259,\#2138286, \#2138307, \#2137603, \#2138296.

\clearpage
\footnotesize
\bibliographystyle{IEEEtran}
\bibliography{refs}

\end{document}